# Spectral Efficiency Optimized Adaptive Transmission for Cognitive Radios in an Interference Channel

Mehrdad Taki and Farshad Lahouti
Wireless Multimedia Communications Laboratory
School of Electrical and Computer Engineering, University of Tehran, Iran

*Abstract*— **In this paper, we consider a primary and a cognitive user transmitting over a wireless fading interference channel. The primary user transmits with a constant power and utilizes an adaptive modulation and coding (AMC) scheme satisfying a bit error rate requirement. We propose a link adaptation scheme to maximize the average spectral efficiency of the cognitive radio, while a minimum required spectral efficiency for the primary user is provisioned. The resulting problem is constrained to also satisfy a bit error rate requirement and a power constraint for the cognitive link. The AMC mode selection and power control at the cognitive transmitter is optimized based on the scaled signal to noise plus interference ratio feedback of both links. The problem is then cast as a nonlinear discrete optimization problem for which a fast and efficient suboptimum solution is presented. We also present a scheme with rate adaption and a constant power. An important characteristic of the proposed schemes is that no negotiation between the users is required. Comparisons with underlay and interweave approaches to cognitive radio with adaptive transmission demonstrate the efficiency of the proposed solutions.**

## I. INTRODUCTION

Cognitive radio, as a promising technology to improve spectrum utilization efficiency, has been the subject of intensive researches recently [1]. In a cognitive radio system, a secondary (cognitive) link is activated along with the primary (licensed) link in a way that it does not disrupt the primary link. There are three well known approaches for the cognitive transmission, namely the interweave, the underlay and the overlay approach [2]. In the interweave approach, the secondary user transmits in spectrum gaps that are not in use by the licensed users. In the underlay approach, the cognitive radio transmits in a manner that its interference at the primary receivers is negligible. In the overlay approach the cognitive radio imposes non-negligible interference at the primary receiver but it makes up the performance degradation in the primary radio with the aid of its non-causal access to the primary users data. An alternative approach for the cognitive transmission is designed in [3] in which the cognitive radio generates non-negligible interference on the licensed receiver but it has a certain "interference budget". The cognitive radio listens to the ARQ of the primary link and estimates its transmission rate. Based on this knowledge the cognitive radio adjusts its rate in a way that a minimum required rate for the primary link is maintained.

Adaptive modulation, coding and power control (AMCP) is shown to have considerable effect on the performance of the wireless systems [4]. It has been also recommended for efficient spectrum utilization in the cognitive radio networks [1]. Several AMCP schemes based on the underlay or interweave approaches for the cognitive radio networks are suggested in [5]-[10] in which the cognitive radio has negligible interference on the primary receivers.

In [11]-[12], power and rate adaption is used in cognitive networks, where the transmitters impose non-negligible interference at the unintended receivers without any compensation. In [11]-[12], the gains of direct and cross links are constant. In [11], the objective is to maximize the cognitive link rate, while an instantaneous rate for the primary link is guaranteed. In [12] the sum of link utility functions and in [13] the sum of links instantaneous rates are subject to maximization..

In [12], the Lagrangian multiplier technique and in others a game theoretic method is employed for the optimization. Therefore, the optimization is accomplished either at a central unit or through negotiation channels between users for a distributed implementation. In general in cognitive radio networks, it is highly desirable that the activity and provisioning of the secondary users do not affect or involve the primary radios.

In this paper, we consider a wireless fading system with a cognitive radio that is concurrently transmitting with a primary user. The latter operates with a constant power and utilizes an adaptive modulation and coding (AMC) scheme satisfying a bit error rate requirement. Two schemes for the cognitive radio operation based on adaptive rate (and power) transmission are proposed to maximize the average spectral efficiency of the cognitive link, while guaranteeing a minimum required average spectral efficiency for the primary link. In both schemes the power constraint and the BER requirement of the cognitive link are provisioned. An important characteristic of the proposed schemes is that no

This work is supported in part by Iran Telecommunications Research Center (ITRC).

negotiation between users is required and the optimization procedure is done at the cognitive radio. The proposed schemes are compared with the underlay and the interweave approaches with adaptive transmission.

## II. System Model

### A. Notations

In this paper, lower case italic letters denote random variables or elements of vectors, e.g. $z$ or $x(i)$. Functions are denoted with lower case letters, e.g. $\mathrm{g}(.)$, and constant parameters are shown with uppercase letters, e.g. N.

### B. System Description and Channel Model

We consider a wireless system in which a primary and a cognitive link are concurrently active as shown in Fig. 1. Each link, $i = 1,2,$ involves a user with a transmitter, $\mathrm{Tx}_i$ wishing to communicate with a corresponding receiver, $\mathrm{Rx}_i$. The links 1 and 2 are respectively considered as the primary and the cognitive links. The channels are assumed discrete time memoryless such that the received signal depends on the transmitted signals as follows:

$$y_i(n) = h_{ii} \times x_i(n) + h_{ji} \times x_j(n) + z_i(n); j = 1,2; i \neq j \quad (1)$$

where $n$ indicates the time index, and $h_{ij}$ denotes the independent and identically distributed channel gain from $\mathrm{Tx}_i$ to $\mathrm{Rx}_j$. The term $h_{ji} \times x_j(n)$ is the interference that is imposed by the unintended transmitter and $z_i(n)$ is additive white Gaussian noise term. We assume frequency flat fading channels with stationary and ergodic time-varying gains. A block-fading model is adopted, where the channel gain remains constant during a block-length (here a codeword), and independently changes from one block to another [14]. The gains of direct and cross links are independent from each other and also independent from the noise.

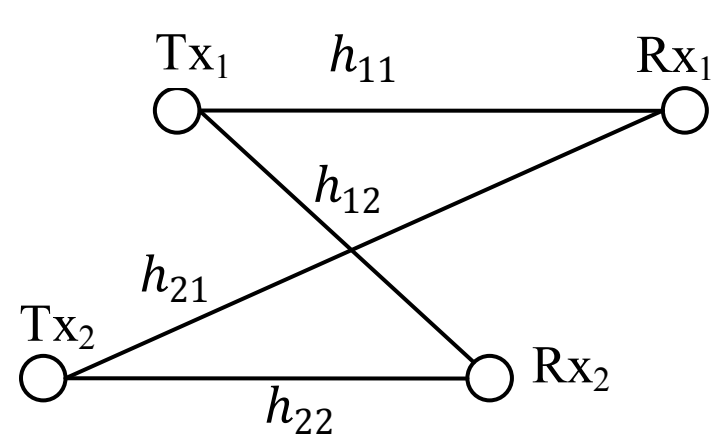

Fig. 1. System configuration: a primary and a cognitive radio

We assume each transmitter uses AMC transmission based on perfect feedback of the signal to noise plus interference power ratio (SNIR) of the direct links. Furthermore, we assume that the primary transmitter operates with a constant power $p_1 = \mathrm{P}_1$, but the cognitive radio may adapt its power $p_2$.

### C. Bit Error Rate Approximation

In general, the interference of the unintended transmitter may be treated as noise at a receiver. Alternatively it may be detected and canceled from the received signal prior to detection of the desired signal. Motivated by the following facts, we take the former approach in this article: (i) In an interference channel, it is shown [15] that when the ratio of interference to the desired signal power is smaller than a threshold (noisy interference), interference should be treated as noise to achieve system capacity. In the current system model, it is assumed that at each receiver the average interference power is weak compared to that of the desired signal; (ii) The detection of primary transmitted signal by the cognitive radio raises certain security problems in practice; (iii) In general, the receiver of the primary link is not necessarily designed to detect and cancel the unintended interference signals of the cognitive radio. We, therefore, as in [11]-[13] assess the performance of each link with its SNIR. The SNIR at $\mathrm{Rx_i}$ is:

$$\gamma_i = \frac{p_i \times s_{ii}}{p_j \times s_{ji} + \mathrm{N}_0} \qquad i,j = 1,2 \quad ; i \neq j \qquad (2)$$

where $\mathrm{N}_0$ is the variance of AWGN and $s_{ji} = \left| h_{ji} \right|^2$.

In an AMC system, there are $\mathrm{N}+1$ transmission modes, each characterized by a modulation and a coding scheme, resulting in a transmission rate, R [4]. The AMC modes are assumed sorted according to their rates, i.e.,

$$0 = \mathrm{R}_0 < \mathrm{R}_1 < \mathrm{R}_2 \ldots < \mathrm{R}_\mathrm{N} \qquad (3)$$

The mode “0” corresponds to no data transmission or outage. The BER performance of the signaling in AWGN channel when the link SNR is $\gamma$, is approximated by a fitting expression as follows [16]:

$$\mathrm{p_e}(\gamma) = \mathrm{f}(\gamma, \mathrm{R}_n) \triangleq \mathrm{E}_n . \exp\left(-\mathrm{Q}_n \times \gamma\right), \quad 0 \leq \gamma \qquad (4)$$

where $\{\mathrm{E}_n , \mathrm{Q}_n\}$ are mode specific constants.

In transmission mode $n$, the minimum required SNIR to guarantee an instantaneous BER smaller than a predetermined value $\mathrm{B}_0$, is given by

$$\mathrm{p_e}(\gamma) \leq \mathrm{B}_0 \Rightarrow \mathrm{f}(\gamma, \mathrm{R}_n) \leq \mathrm{B}_0 \Rightarrow \gamma \geq \mathrm{g}_{\mathrm{B}_0}(\mathrm{R}_n), \qquad (5)$$

where the function $\mathrm{g}_{\mathrm{B}_0}(\mathrm{R}_n)$ is defined as:

$$\mathrm{g}_{\mathrm{B}_0}(\mathrm{R}_n) \triangleq -\frac{1}{\mathrm{Q}_n} \times \ln\left(\frac{\mathrm{B}_0}{\mathrm{E}_n}\right), \ \mathrm{B}_0 \leq \mathrm{E}_n \qquad (6)$$

## III. Link Adaptation for Cognitive Transmission

In this section, we propose adaptive transmission schemes for the introduced system to maximize the average spectral efficiency of the cognitive link, while satisfying a minimum required average spectral efficiency for the primary link and the power constraint on the cognitive transmitter. It is assumed that the application requires a maximum BER of $\mathrm{B}_1$ for the primary link and $\mathrm{B}_2$ for the cognitive link. In both schemes the primary user transmits with a constant power $\mathrm{P}_1$ and utilizes an AMC scheme to satisfy its BER constraint. We propose two schemes for the cognitive transmission; constant power and adaptive power link adaptation. In the both approaches adaptation is based on link SNIRs that are fed back to the transmitters.

### A. Constant Power Link Adaptation Scheme

In this scheme, the cognitive user transmits with a constant power $\mathrm{P}_2$ that is selected in a way that the primary user can achieve its required average spectral efficiency, $\overline{\mathrm{K}_1}$. The link AMC rates are denoted by $k_1(\gamma_1), k_2(\gamma_2)$ for the primary and the cognitive link, respectively. Both radios adapt their AMC rates based on their own link SNIR to satisfy their BER requirements. If $\overline{\mathrm{P}}_2$ is the maximum power constraint of the cognitive transmitter, the link adaption problem is formulated

as:

$$\max_{P_2(\gamma_2)} \int_0^\infty k_2(\gamma_2).\mathrm{pr}_2(\gamma_2)\, d\gamma_2 \qquad \text{subject to:} \qquad (7)$$

$$\begin{cases} \mathrm{C1:} \int_0^\infty k_1(\gamma_1).\mathrm{pr}_1(\gamma_1)\, d\gamma_1 \ge \overline{\mathrm{K}_1} \\ \mathrm{C2:}\ \mathrm{P}_2 \le \overline{\mathrm{P}}_2 \\ \mathrm{C3:}\ \mathrm{f}(\gamma_1, k_1(\gamma_1)) \le \mathrm{B}_1 \\ \mathrm{C4:}\ \mathrm{f}(\gamma_2, k_2(\gamma_2)) \le \mathrm{B}_2 \end{cases}$$

In (7), $\mathrm{pr}_1(\gamma_1)$ and $\mathrm{pr}_2(\gamma_2)$ are the probability density functions of SNIRs. Note that both $\gamma_1$ and $\gamma_2$ are functions of $\mathrm{P}_2$.

As in [4], the range of SNIR of the transmitter $m \in \{1,2\}$ is divided into $\mathrm{N}+1$ non-overlapping consecutive intervals, where interval $n$ is denoted by $[\nu_{m,n}, \nu_{m,n+1})$ for $0 \le n \le \mathrm{N}$ and $\nu_{m,0} = 0, \nu_{m,\mathrm{N}+1} = \infty$. If the SNIR falls in the interval $n$, the AMC transmission mode $n$ with rate $\mathrm{R}_n$ is selected. The average spectral efficiency of each link is computed as:

$$k_m^{avg} = \sum_{i=1}^{\mathrm{N}} \mathrm{R}_i \times \mathrm{pr}\{\nu_{m,i} \le \gamma_m < \nu_{m,i+1}\}, m = 1,2. \qquad (8)$$

Given that the power constraint is satisfied, achieving the maximum possible average spectral efficiency of a link, while satisfying the BER constraint results in:

$$\nu_{m,n} = \min_{\gamma_m} \gamma_m \qquad \text{subject to: } \gamma_m \ge \mathrm{g}_{\mathrm{B}_m}(\mathrm{R}_n), m = 1,2. \quad (9)$$

Therefore, $\nu_{m,n} = \mathrm{g}_{\mathrm{B}_m}(\mathrm{R}_n)$. Using above equations the optimization problem in (7) is restated as follows.

$$\max_{\mathrm{P}_2} k_2^{avg} \quad \text{subject to: } \begin{cases} \mathrm{C1:} k_1^{avg} \ge \overline{\mathrm{K}_1} \\ \mathrm{C2:}\ \mathrm{P}_2 \le \overline{\mathrm{P}}_2 \end{cases} \qquad (10)$$

Note that increasing $\mathrm{P}_2$, increases $k_2^{avg}$ and decreases $k_1^{avg}$. The solution for (10) is obtained by increasing $\mathrm{P}_2$ from zero up to a point where either C1 or C2 is satisfied with equality.

### *B. Variable Power Link Adaptation Scheme*

In the second scheme, the cognitive user adapts its power $p_2(\gamma_1, \gamma_2)$ and rate $k_2(\gamma_1, \gamma_2)$ to optimize the link utilization. It is aware of the primary link SNIR in addition to that of its own, e.g. it can listen to the CSI feedback of both links. In the following subsections, we first setup the link adaptation problem and next reformulate it in a form for which an effective solution is presented. The solution sets the values of the variables $k_2(\gamma_1, \gamma_2)$, $p_2(\gamma_1, \gamma_2)$ and $k_1(\gamma_1)$.

#### *1) Problem Setup*

The described link adaptation problem is formulated as follows:

$$\max_{k_2(.), p_2(.)} \int_0^\infty \int_0^\infty k_2(\gamma_1, \gamma_2)\mathrm{pr}(\gamma_1, \gamma_2) d\gamma_1\, d\gamma_2, \text{ subject to:}$$

$$\begin{cases} \mathrm{C1:} \int_0^\infty \int_0^\infty k_1(\gamma_1)\mathrm{pr}(\gamma_1) d\gamma_1 \ge \overline{\mathrm{K}_1} \\ \mathrm{C2:} \int_0^\infty \int_0^\infty p_2(\gamma_1, \gamma_2)\mathrm{pr}(\gamma_1, \gamma_2) d\gamma_1\, d\gamma_2 \le \overline{\mathrm{P}}_2 \\ \mathrm{C3:}\ \mathrm{f}(\gamma_1, k_1(\gamma_1)) \le \mathrm{B}_1 \\ \mathrm{C4:}\ \mathrm{f}(\gamma_2, k_2(\gamma_1, \gamma_2)) \le \mathrm{B}_2 \end{cases} \qquad (11)$$

where $\mathrm{pr}(\gamma_1, \gamma_2)$ is a joint probability density function of SNIRs. Due to the BER constraint on the cognitive link, (C4 in Eq. (11)), the variable $p_2(\gamma_1, \gamma_2)$ depends on $k_2(\gamma_1, \gamma_2)$; and hence when necessary it is denoted by $p_2\left(\gamma_1, \gamma_2,\ k_2(\gamma_1, \gamma_2)\right)$. It is clear that the SNIR of the primary link is affected by $p_2(.)$, therefore, the rate of the primary link, is indirectly a function of both $\gamma_1$ and $\gamma_2$, and is denoted by $k_1(\gamma_1, \gamma_2)$. As $p_2(.)$ is a function of $k_2(\gamma_1, \gamma_2)$, we represent the rate of the primary link by $k_1(\gamma_1, \gamma_2,\ k_2(\gamma_1, \gamma_2))$ when the emphasis is necessary. Using these relations the optimization problem is restated as:

$$\max_{k_2(.)} \int_0^\infty \int_0^\infty k_2(\gamma_1, \gamma_2)\mathrm{pr}(\gamma_1, \gamma_2) d\gamma_1\, d\gamma_2 \text{ subject to:} \qquad (12)$$

$$\begin{cases} \mathrm{C1:} \int_0^\infty \int_0^\infty k_1\left(\gamma_1, \gamma_2,\ k_2(\gamma_1, \gamma_2)\right) \mathrm{pr}(\gamma_1, \gamma_2) d\gamma_1\, d\gamma_2 \ge \overline{\mathrm{K}_1} \\ \mathrm{C2:} \int_0^\infty \int_0^\infty p_2\left(\gamma_1, \gamma_2,\ k_2(\gamma_1, \gamma_2)\right) \mathrm{pr}(\gamma_1, \gamma_2) d\gamma_1\, d\gamma_2 \le \overline{\mathrm{P}}_2 \\ \mathrm{C3:}\ \mathrm{f}\left(\gamma_1, k_1(\gamma_1, \gamma_2, k_2(\gamma_1, \gamma_2))\right) \le \mathrm{B}_1 \\ \mathrm{C4:}\ \mathrm{f}(\gamma_2, k_2(\gamma_1, \gamma_2)) \le \mathrm{B}_2 \end{cases}$$

As evident, the resulting optimization problem is complex and cannot be directly solved. In the following, we reformulate it in a more tractable form for which a solution is presented.

#### *2) Problem Formulation*

In this section, we reformulate the desired optimization problem in Eq. (12), first by re-examining the effect of interference on link adaptation, and subsequently by exploiting the discrete nature of AMC transmission rates.

Here, we assume that the additive thermal noise at the receivers is in general negligible, when its power is compared to that of the interference signal. This assumption during the design may result in a violation of the constraints C3 and C4 in Eq. (12). To address this issue, as analyzed in [17], one may consider a tighter BER constraint during the design than that required by the application. The received SNIRs is then given by

$$\begin{cases} p_2 > 0 \quad \Rightarrow \begin{cases} \gamma_1 = \frac{p_1 s_{11}}{p_2 s_{21}} = p_1\alpha/p_2 \\ \gamma_2 = \frac{p_2 s_{22}}{p_1 s_{12}} = p_2\beta/p_1 \end{cases}, \\ p_2 = 0 \quad \Rightarrow \begin{cases} \gamma_1 = \frac{p_1 s_{22}}{\mathrm{N}_0} \\ \gamma_2 = 0 \end{cases} \end{cases} \qquad (13)$$

where $\alpha$ and $\beta$ are the scaled SNIRs of the primary and cognitive links. It is clear from definitions that $\alpha$ and $\beta$ are independent random variables.

We express $k_1(.)$ and $p_2(.)$ in terms of $k_2(.)$. Satisfying C4 in Eq. (12) with equality, noting Eq. (13), the power of the cognitive transmitter is given by

$$p_2(.) = \mathrm{P}_1\, \mathrm{g}_{\mathrm{B}_2}(k_2(.))/\beta, \quad k_2(.) > 0. \qquad (14)$$

This transmission power results in the following SNIR at the primary receiver

$$\gamma_1 = \alpha\beta/\mathrm{g}_{\mathrm{B}_2}(k_2(.)). \qquad (15)$$

The primary user selects the maximum rate from AMC table that satisfies its BER requirement, i.e.,

$$k_1(.) = \arg\max_{\mathrm{R}} \mathrm{g}_{\mathrm{B}_1}(\mathrm{R}) \text{ subject to: } \mathrm{g}_{\mathrm{B}_1}(\mathrm{R}) \le \gamma_1,\ k_2(.) > 0, \qquad (16)$$

Noting Eq.'s (13) and (16), when the cognitive radio transmission power is $p_2(.) > \mathrm{P}_1\alpha/\mathrm{g}_{\mathrm{B}_2}(\mathrm{R}_1)$, the primary link BER requirement, even with its lowest AMC rate, is violated and hence an outage occurs.

According to Eq. (13), when $k_2(.) = 0$ and hence $p_2(.) = 0$, the primary user selects its rate based on the SNR of its link and its average rate given the BER constraint is given by

$$k_1(\alpha, \beta, k_2(\alpha, \beta) = 0) = \sum_{i=1}^{\mathrm{N}} \mathrm{R}_i \times \mathrm{pr}\left\{\nu_{1,i} \le \frac{\mathrm{P}_1 s_{11}}{\mathrm{N}_0} < \nu_{1,i+1} \middle| \alpha, \beta\right\}, \qquad (17)$$

where $\nu_{1,i}$ is computed in (9).

The optimization problem in Eq. (12) is now restated as follows based on $p_2(\alpha,\beta,k_2(\alpha,\beta))$ and $k_1(\alpha,\beta,k_2(\alpha,\beta))$ computed in Eq.'s (14), (16) and (17).

$$\max_{k_2(\alpha,\beta)} \int_0^\infty \int_0^\infty k_2(\alpha,\beta)\mathrm{pr}(\alpha)\mathrm{pr}(\beta)d\alpha d\beta \quad \text{subject to:}$$

$$\begin{cases} \text{C1: } \int_0^\infty \int_0^\infty k_1(\alpha,\beta,k_2(\alpha,\beta))\mathrm{pr}(\alpha)\mathrm{pr}(\beta)d\alpha d\beta \geq \overline{\mathrm{K}_1} \\ \text{C2: } \int_0^\infty \int_0^\infty \mathrm{P}_1 \frac{\mathrm{g}_{\mathrm{B}_0}(k_2(\alpha,\beta))}{\beta}\mathrm{pr}(\alpha)\mathrm{pr}(\beta)d\alpha d\beta \leq \overline{\mathrm{P}_2} \end{cases} \tag{18}$$

The BER constraints, C3 and C4, in Eq. (12) are now considered in the power and rate assignments in Eq. (18).

We next consider the discrete nature of the AMC scheme to convert the problem to a discrete and manageable form. We present several definitions toward this conversion.

*Definition: Permissible rate pairs set (rate set)*

For each point in the $\alpha-\beta$ plane, based on Eq. (16), there are certain rate pairs that are valid for the primary and cognitive links. This set of rate pairs is referred to as the permissible rate pairs set or simply rate set of that point.

*Definition: Common rate set regions*

As the primary and cognitive rates in Eq. (16) are discrete variables, certain regions partitioning the $\alpha-\beta$ plane are formed, in which the corresponding rate set remains constant. In general, there are $\mathrm{M}=\mathrm{N}^2+1$ such regions that are referred to as common rate set regions. These are analogous to the non-overlapping intervals partitioning the SNR range in the single link AMC design. The common rate set region $i$ is defined as:

$$\mathrm{Z}_{i-1} \leq \alpha\beta \leq \mathrm{Z}_i \;, \quad 1 \leq i \leq \mathrm{M}+1 \tag{19}$$

where $Z_i \in \{\mathrm{g}_{\mathrm{B}_1}(\mathrm{R}_n)\times \mathrm{g}_{\mathrm{B}_2}(\mathrm{R}_m), 1\leq m,n\leq \mathrm{N}\}$, for $1\leq i\leq$ M and "$Z_i$"s are sorted in ascending order $\mathrm{Z}_1\leq \mathrm{Z}_2\leq\cdots\leq \mathrm{Z}_\mathrm{M}$ and $\mathrm{Z}_0=0$ and $\mathrm{Z}_{\mathrm{M}+1}=\infty$. In Fig. 2, for a given set of AMC modes, described in section V, the common rate set regions are identified as the area between two consecutive solid lines. These boundary lines correspond to $\alpha\beta=\mathrm{Z}_i, 0\leq i\leq \mathrm{M}+1$.

*Definition: Common rate regions*

The common rate regions are areas in the $\alpha-\beta$ plane, each of which belongs to one common rate set region and assigned the same rate pair. These areas are indentified by the boundary lines described above and sufficient auxiliary boundary lines as follows:

$$\beta/\alpha=\mathrm{W}_j,\; j=0,..,\mathrm{L} \tag{20}$$

$$\alpha\beta=\mathrm{Q}_h,\; h=1,\ldots,\mathrm{C}-\mathrm{M} \tag{21}$$

where $\mathrm{W}_j$'s and $\mathrm{Q}_i$'s are constants and $\mathrm{W}_0=0$, $\mathrm{W}_\mathrm{L}=\infty$. These lines are depicted as dashed lines in Fig. 2.

Using the mentioned boundaries the $\alpha-\beta$ plane is divided into $\Upsilon=\mathrm{L}\times\mathrm{C}$ common rate regions, that are denoted by $reg(i), 1\leq i\leq\Upsilon$, and are simply called regions in the rest of the paper. As the rates in a given region are fixed, when the scaled SNIRs fall into $reg(i)$, the rates assigned to the cognitive and primary links (two dependent variables) are denoted by $k_2(i)$ and $k_1(i,\ k_2(i))$ (or simply $k_1(i)$).

With these definitions the average spectral efficiency of the primary and cognitive links ($k_1^{avg}, k_2^{avg}$) are computed as follows:

$$k_2^{avg}=\sum_{i=1}^{\Upsilon} k_2(i)\times \mathrm{pr}(i) \tag{22}$$

$$k_1^{avg}=\sum_{i=1}^{\Upsilon} k_1(i,k_2(i))\times \mathrm{pr}(i) \tag{23}$$

where $\mathrm{pr}(i)$ is the probability that the scaled SNIRs fall into $reg(i)$. The relation between $k_1(.)$ and $k_2(.)$ is as in Eq. (16).

*Remark:* If the cognitive link is inactive $k_2(i)=0$, based on Eq. (14), the average rate of the primary link when SNIRs are in $reg(i)$ is:

$$k_2(i)=0 \quad\Rightarrow\quad k_1(i)=\sum_{i=1}^{\mathrm{N}} \mathrm{R_i}\times \mathrm{pr}\left\{\nu_{1,i}\leq \frac{\mathrm{P}_1 s_{11}}{\mathrm{N}_0}< \nu_{1,i+1}\,\middle|\, \alpha,\beta\in reg(i)\right\}. \tag{24}$$

The average power that the cognitive radio uses to transmit with the rate $k_2(i)$ in $reg(i)$, noting Eq. (14), is given by

$$power(i)=\mathrm{P}_1\iint_{reg(i)} \frac{\mathrm{g}_{\mathrm{B}_2}(k_2(i))}{\beta}\mathrm{p}(\alpha)\mathrm{p}(\beta)\,d\alpha\,d\beta \tag{25}$$

The normalized power in $reg(i)$ is defined as

$$p(i)\overset{\text{def}}{=} power(i)/\mathrm{g}_{\mathrm{B}_2}(k_2(i))\times \mathrm{pr}(i) \tag{26}$$

The normalized average power that is used in cognitive transmitter is:

$$p_2^{avg}=\sum_{i=1}^{\Upsilon} \mathrm{g}_{\mathrm{B}_2}(k_2(i))\,p(i)\,\mathrm{pr}(i) \tag{27}$$

It is clear that when the number of regions goes to infinity, the summations in equations (22), (23) and (27) approach their corresponding values in Eq. (18). The desired optimization problem can be restated as follows

$$\max_{k_2(i),1\leq i\leq\Upsilon} k_2^{avg}=\sum_{i=1}^{\Upsilon} k_2(i)\,\mathrm{pr}(i) \quad \text{subject to:}$$

$$\begin{cases} \text{C1:} k_1^{avg}=\sum_{i=1}^{\Upsilon} k_1(i,k_2(i))\,\mathrm{pr}(i)\geq \overline{\mathrm{K}_1} \\ \text{C2:} p_2^{avg}=\sum_{i=1}^{\Upsilon} \mathrm{g}_{\mathrm{B}_2}(k_2(i))\,p(i)\,\mathrm{pr}(i)\leq \overline{\mathrm{P}_2} \end{cases} \tag{28}$$

### 3) Problem Solution

In general, one may consider solving the problem in Eq. (28) by nonlinear programming methods [18]. The optimality and convergence of these iterative algorithms relies on the proper definition of the gradient function and the second-order differentiation of the corresponding continuous forms of objective function. However, the discrete function, $k_1(i,k_2(i))$, does not have a closed form expression that is differentiable. In the followings, we propose a fast and simple alternative iterative algorithm that is inspired by gradient methods and enabled with a proper definition of gradient function and selection of the initial point. In [17], the optimality of the algorithm in certain cases is proved. It is also shown that in general the difference between the obtained rates and optimal values is of the order of probability of one region that can be very small. To describe the proposed algorithm, we first present several definitions:

*Definitions: Decision Variables*

The variable $d_1(i)$ is defined as follows:

$$d_1(i)\triangleq \begin{cases} 0, & k_1(\mathrm{R}_m,i)=\mathrm{R}_\mathrm{N} \\ -\frac{\Delta k_1^{avg}}{\Delta k_2^{avg}}=-\frac{k_1(\mathrm{R}_m,i)-k_1(\mathrm{R}_{m-n},i)}{\mathrm{R}_m-\mathrm{R}_{m-n}}, & k_1(\mathrm{R}_m,i)<\mathrm{R}_\mathrm{N} \end{cases} \tag{29}$$

where $1\leq i\leq\Upsilon$; $n>0$, $\mathrm{R}_{m-n}\geq 0$; $k_2(i)=\mathrm{R}_m$ and $n$ is computed as follows:

$$n=\arg\min_x -\big(k_1(\mathrm{R}_m,i)-k_1(\mathrm{R}_{m-x},i)\big) \text{ subject to: } -\big(k_1(\mathrm{R}_m,i)-k_1(\mathrm{R}_{m-x},i)\big)>0 \tag{30}$$

In the Eq. (29), $\Delta k_1^{avg}/\Delta k_2^{avg}=0$ is obtained based on the fact that when $k_1(\mathrm{R}_m,i)$ is at maximum $\mathrm{R}_\mathrm{N}$, decreasing the rate of the cognitive link has no effect on the rate of the primary link.

The variable $d_2(i)$ is defined as:

$$d_2(i,n) \triangleq \frac{\Delta p_2^{avg}}{\Delta k_2^{avg}} = \frac{\left(g_{B0}(R_m) - g_{B0}(R_{m-n})\right) \times p(i)}{R_m - R_{m-n}} \quad (31)$$

where $1 \leq i \leq \Upsilon$; $n > 0$, $R_{m-n} \geq 0$; $k_2(i) = R_m$.

The variable $d_3(i)$ is defined as:

$$d_3(i) \triangleq \begin{cases} 0, & k_1(R_m, i) = R_N \\ d_2(i,n), & k_1(R_m, i) < R_N \end{cases} \quad (32)$$

where $n$ is obtained from Eq. (30).

In summary, the proposed algorithm first assigns the maximum rate $R_N$ to the cognitive link in all regions. Next, it reduces the assigned rate in some appropriate regions to satisfy the constraints. The steps of the algorithm are detailed below.

**Proposed Algorithm:**

1. Set $k_2(i) = R_N, 1 \leq i \leq \Upsilon$.
2. Compute $k_1(i), 1 \leq i \leq \Upsilon$ based on Eq. (16) and $k_1^{avg}$ and $p_2^{avg}$ based on Eq.'s (23) and (27).
3. If none of the constraints are satisfied ($k_1^{avg} \leq \overline{K_1}$ and $p_2^{avg} \geq \overline{P_2}$) go to step 4. Else if only the constraint on average spectral efficiency of link 1 is not satisfied ($k_1^{avg} \leq \overline{K_1}$ and $p_2^{avg} \leq \overline{P_2}$) go to step 8. Else if only the constraint on the average power of the link 2 is not satisfied ($k_1^{avg} \geq \overline{K_1}$ and $p_2^{avg} \geq \overline{P_2}$) go to step 12. Otherwise ($k_1^{avg} \geq \overline{K_1}$ and $p_2^{avg} \leq \overline{P_2}$) go to step 16.
4. Compute values of $d_3(i), 1 \leq i \leq \Upsilon$ based on Eq. (32).
5. Find $i_m = \arg\max_i d_3(i)$. If $k_2(i_m) = 0$, go to step 17 else if $k_2(i_m) = R_m > 0$, set $k_2(i_m) = R_{m-n}$, where $n$ is given in Eq. (30).
6. Update $k_1^{avg}$ and $p_2^{avg}$ using the next equations.
   $$k_1^{avg} \leftarrow k_1^{avg} - \left(k_1(R_m, i) - k_1(R_{m-n}, i)\right) \times \mathrm{pr}(i) \quad (33)$$
   $$p_2^{avg} \leftarrow p_2^{avg} - (g_{B0}(R_m) - g_{B0}(R_{m-1})) \times \mathrm{pr}(i) \times p(i) \quad (34)$$
   and $d_3(i_m)$ based on Eq. (32).
7. If $k_1^{avg} \leq \overline{K_1}$ and $p_2^{avg} \geq \overline{P_2}$, go to step 5 else go to step 3.
8. Compute $d_1(i)$, $1 \leq i \leq \Upsilon$, using Eq. (29).
9. Find $i_m = \arg\max_i d_1(i)$. If $k_2(i_m) = 0$, go to step 17 else if $k_2(i_m) = R_m > 0$, set $k_2(i_m) = R_{m-n}$, where $n$ is given in Eq. (30).
10. Update $k_1^{avg}$ using Eq. (33) and $d_1(i_m)$ based on Eq. (29).
11. If $k_1^{avg} \leq \overline{K_1}$ go to step 9, else go to step 16.
12. Compute $d_2(i,1)$,$1 \leq i \leq \Upsilon$, based on Eq. (31).
13. Find $i_m = \arg\max_i d_2(i,1)$. If $k_2(i_m) = 0$, go to step 17 else if $k_2(i_m) = R_m > 0$, set $k_2(i_m) = R_{m-1}$;
14. Update $p_2^{avg}$ using Eq. (34) and $d_2(i_m)$ using Eq. (31).
15. If $p_2^{avg} \geq \overline{P_2}$ go to step 13.
16. End. The desired design variables $k_2(i)$ are obtained.
17. The constraints can not be provisioned.

In [17], it is shown that the algorithm complexity is $O(\Upsilon)$, per iteration and the maximum number of iterations to convergence is $O(\Upsilon)$. Therefore, the total complexity of the algorithm is $O(\Upsilon^2)$ in the worst case. It is noteworthy that the complexity of an exhaustive search is $O(N^\Upsilon)$.

## IV. Comparison With Underlay And Interweave Approaches

For comparison, we consider the underlay and the interweave approaches for cognitive radio transmission within the system model described in section II.B. In the underlay approach, the cognitive radio adapts its power, on a per block basis, to utilize its link in a way that the power of its interference at the primary receiver is smaller than a threshold, $P_{th}$ [7]. Different values of $P_{th}$ result in different average spectral efficiencies for the primary and cognitive links. In the interweave approach, the cognitive radio transmits with an AMCP scheme, when the primary link is inactive [6]. In this case, the optimal scenario is when the two transmitters share the resources in a coordinated manner, e.g., using TDMA. Further details on the performance analysis of these schemes are presented in [17].

## V. Performance Evaluation

We use the AMC transmission modes of the IEEE 802.11a standard [19] for performance evaluation. There are eight modes set up based on different convolutionally coded QAM modulations with rates $R_i \in \{0,0.5,0.75,1,1.5,2,3,4\}$. The model parameters according to Eq. (4) are derived in [20].

In Fig. 3, the average spectral efficiency of the cognitive link obtained using the proposed adaptive transmission scheme is depicted as a function of the minimum required average spectral efficiency of the primary link. The results are for a weak interference channel model, described in section II, with the parameters $\overline{s_{11}} = \overline{s_{22}} = 1$, $\overline{s_{12}} = \overline{s_{21}} = 0.05$, and are presented for different number of regions, $\Upsilon$, $(V_0 = 100)$. It is clear that choosing $\Upsilon = 100$, provides accurate results and any further increase leads to only negligible performance improvement. It is also evident that the proposed algorithm for cognitive link adaptation performs very closely with a computationally complex solution based on genetic algorithm.

In Fig. 4, the performance of the system for different approaches of cognitive radio is depicted for the same average power and BER constraints. Two interesting observations are made: (i) The proposed variable power schemes outperform the underlay and interweave approaches and the performance of the constant power scheme is close to the underlay approach; (ii) The presented optimized power link adaptation, when compared to the constant power scheme, provides considerable performance improvement.

In Fig. 5, the case with a large scale path-loss model is considered, i.e., $s_{ij} = 1/d_{ij}^{E}$, where $d_{ij}$ is the distance between $Tx_i$ and $Rx_j$, and E is the path loss exponent (here $E = 3$). The transceivers are positioned on the vertices of a normalized rectangle, i.e., $d_{11} = d_{22} = 1$ and $d_{12} = d_{21} = \sqrt{1 + d^2}$, where $d$ is the distance between the transmitters (receivers). As expected, increasing $d$ reduces the level of interference and hence improves the performance.

## VI. Conclusions

We proposed a new cognitive radio transmission approach for a primary and a cognitive user transmitting over a wireless fading interference channel. In the presented scheme, the cognitive radio utilizing primary and cognitive link SNIRs adapts its link to maximize its spectral efficiency, while considering a minimum required average spectral efficiency for the primary link. Comparisons with the adaptive underlay and interweave approaches to cognitive radio demonstrate a considerable improvement in the system efficiency.

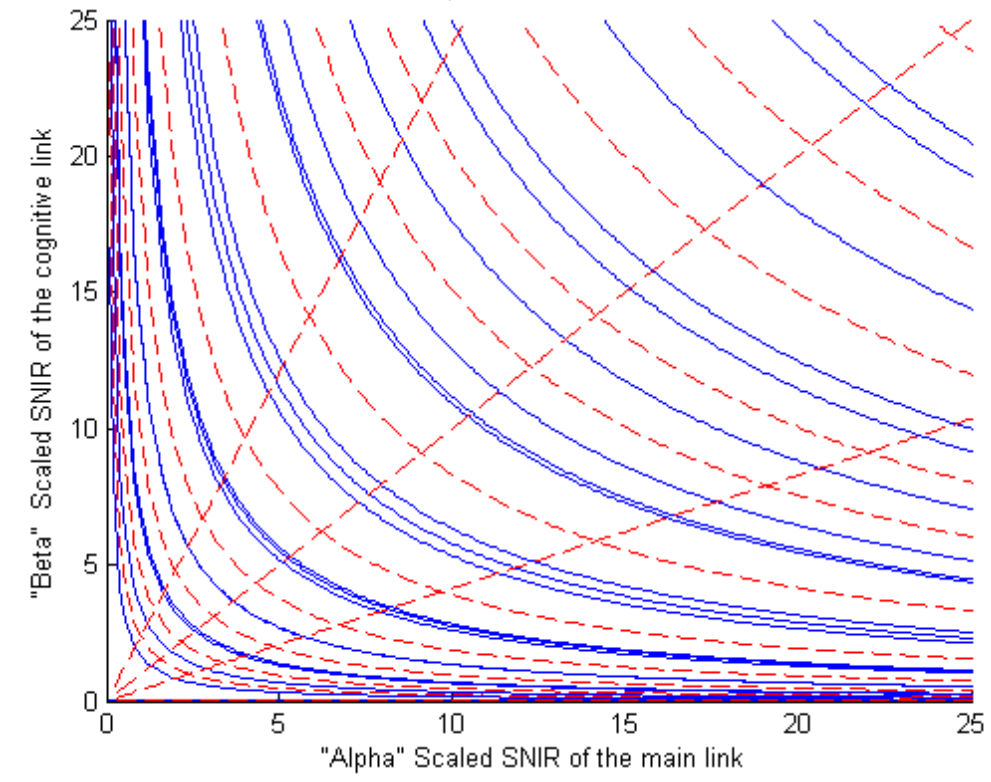

Fig. 2. Partitioning of $\alpha - \beta$ plane into common rate (set) regions.

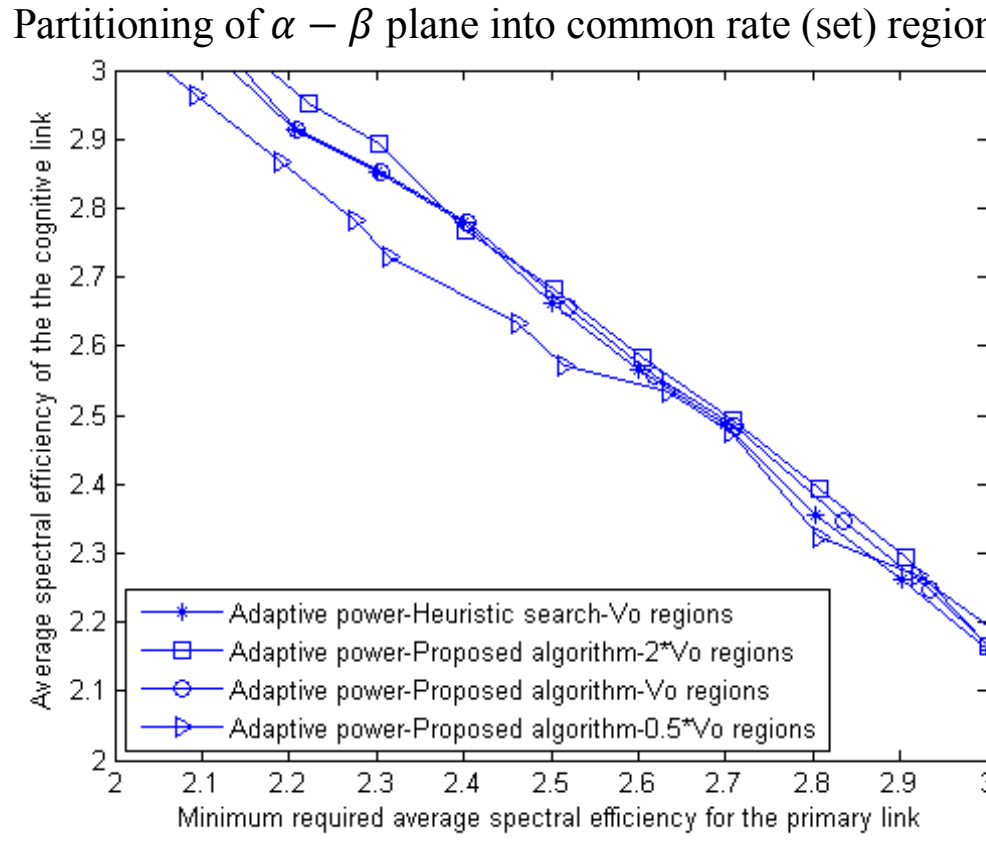

Fig. 3. Spectral efficiency of cognitive vs. primary link; Effect of number of regions and optimization method

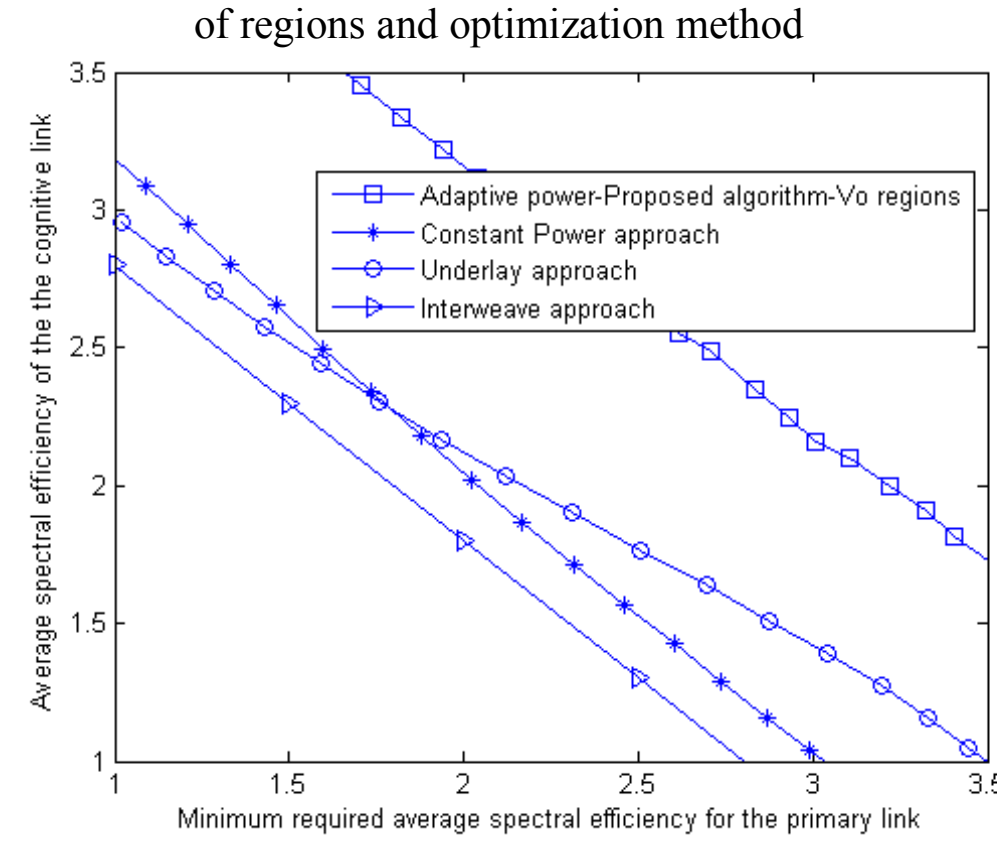

Fig. 4. Spectral efficiency of cognitive vs. primary link; Different approaches

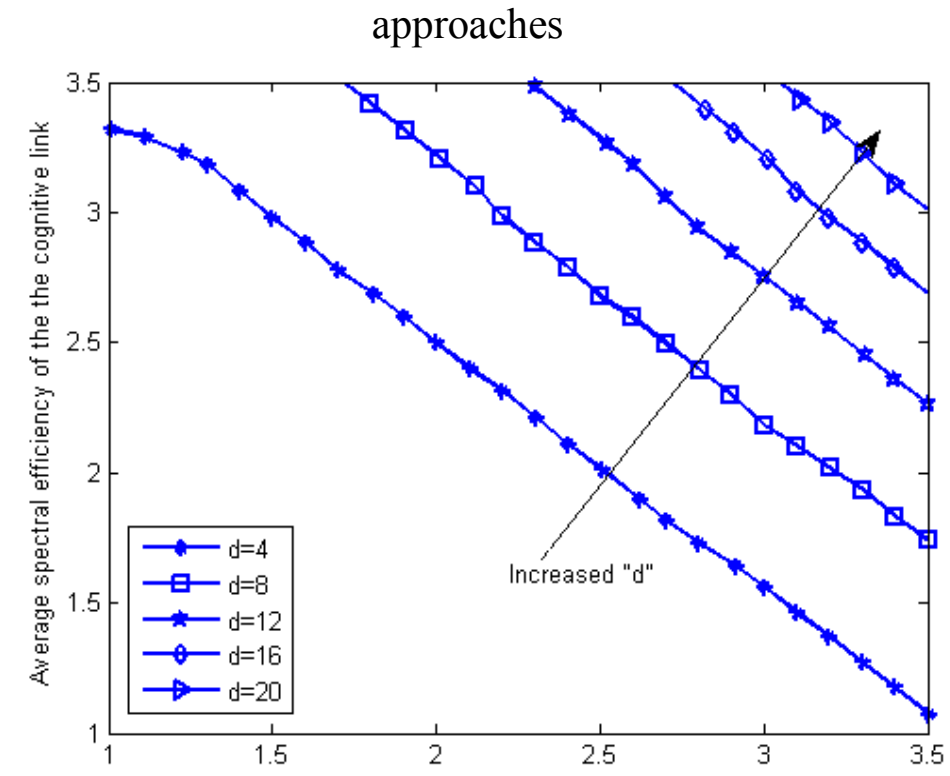

Fig. 5. Spectral efficiency of cognitive vs. primary link for various $d$’s